# A Protocol to Exposure Path Analysis for Multiple Stressors Associated with Cardiovascular Disease Risk: A Novel Approach Using NHANES Data

Jiangling Liu[a,1], Ya Liu[a,1], Banyun Zheng[a,1], Longjian Liu[b,2], Heqing Shen[a,2]

[1]Jiangling Liu, Ya Liu and Banyun Zheng contributed equally to this work.
[2]To whom correspondence may be addressed.

Author affiliations: [a]State Key Laboratory of Vaccines for Infectious Diseases, Xiang An Biomedicine Laboratory & State Key Laboratory of Molecular Vaccinology and Molecular Diagnostics, School of Public Health, Xiamen University, Xiamen, 361102, China; [b]Department of Epidemiology and Biostatistics, Dornsife School of Public Health, Drexel University, Philadelphia, PA, 19104, USA.

Corresponding authors:
Heqing Shen, School of Public Health, Xiamen University, 4221-117 Xiang'an South Road, Xiamen, 361102, China; Email: hqshen@xmu.edu.cn.
ORCID of the author: https://orcid.org/0000-0002-6544-4919.
Longjian Liu: Drexel University, Longjian.Liu@Drexel.edu
Jiangling Liu: Xiamen University, liujiangling@stu.xmu.edu.cn
Ya Liu: Xiamen University, 32620221150850@stu.xmu.edu.cn
Banyun Zheng: Xiamen University, zhengbanyun@stu.xmu.edu.cn

**Declaration of competing financial interests (CFI):** The authors of this manuscript hereby declare that there are no financial or personal relationships with other people or organizations that could inappropriately influence or bias the content of this paper. There are no competing financial interests to disclose.

**Abstract**

**Background:** Multiple medical and non-medical stressors, along with the complicity of their exposure pathways, have posted significant challenges to the epidemiological interpretation of the non-communicable diseases, including cardiovascular disease (CVD).

**Objective:** To develop a protocol for deconstructing the complex exposure pathways linking various stressors to adverse outcomes and to elucidate the sequential determinants contributing to CVD risk in depth.

**Methods:** In this study, we developed a Path-Lasso approach, rooted in Adaptive Lasso regression, to construct the network and paths to interpret the determinants of CVD in an in-depth way by using data from the National Health and Nutrition Examination Survey (NHANES). Univariate logistic regression was initially employed to screen out all potential factors of influencing CVD. Then a programmed approach, using Path-Lasso technique, stratified covariates and established a causal network to predict CVD risk.

**Results:** Age, smoking and waist circumference were identified as the most significant predictors of CVD risk. Other factors, such as race, marital status, physical activity, cadmium exposure and diabetes acted as the intermediary or proximal variables. All these stressors (or nodes) formed the network with paths (or edges to link the CVD), in which the latent layer variables that causally associate to the outcome are linearly formed by the stressors in each layer.

**Discussion:** The Path-Lasso approach revealed the epidemiological pathways, linking covariates to CVD risk, which is instrumental in elucidating the inter-covariate transitions of their predication to the outcome, and providing the hierarchal network for foundation of the assessment of CVD risk and the beyond.

# 1 Introduction

The medical and non-medical stressors that predict the risk of a specific disease or adverse health outcome is usually multivariable and interactive. From a broad vision, a stressor can be any factor to which a human is exposed, from the point of contact between the stressor and receptor, moving inward into the human body and outward to the general environment, including the ecosphere. This means that the exposure pathways can be traced to the ecologic level, community level, or the individual level[1]. Since a stressor can be any physical, chemical and biological entity or socioeconomic condition that induces an adverse response, the World Health Organization (WHO) defines social determinants of health (SDOH) as the factors influencing health outcomes. All determents together form the source-to-outcome continuum for human exposure, which can also be categorized at multiple levels, including different groups or within an individual from the internal exposure to the target-sites. The multiple stressors are often hierarchically organized based on their common action on the defined outcome. For example, alcohol consumption are generally unhealthy that can work as the negative factor, increasing the risk of over 60 diseases[2]. Genome-wide association study (GWAS) has identified that certain single-nucleotide polymorphisms (SNPs), such as ADH1B (rs1229984), KLB (rs13130794), BTF3P13 (rs144198753), GCKR (rs1260326), SLC39A8 (rs13107325) and DRD2 (rs11214609), that are related to the excessive consumption of alcohol. It is also well-known that the distribution of SNPs varies by race and ethnicity. For example, ADH1B and other ADH and ALDH variants associated with alcohol consumption occur at low frequencies among populations of European ancestry but are more common in populations of East Asian ancestry[3]. Given exposure to alcohol can directly impact human health, the internal factors of SNPs can directly modify the alcohol risk that associate the diseases via excessive alcohol consumption, and the racial identity can modify the distribution of SNPs. A simplified pathway can be drawn from race to SNPs to disease in a population. Meanwhile, alcohol consumption can be influenced by many other factors, such as personal habits, economic status, traditional culture, and so on.

In the context of any stressor that increases the risk of adverse outcome, distal stressors may hierarchically regulate proximal ones. This means that information, such as variable variation, can flow from the upstream stressor to downstream stressors along these pathways. Mapping these pathways would be useful for public health prevention. Basically, the follow of information from stressor to stressor toward a health outcome indicates some redundant. In other words, downstream variable transit the same information, at least partially the same, to the outcome as the upstream variable. Therefore, the redundant analysis of the multiple variables' association to the same outcome will be meaningful to address the potential pathways. Given the multiple stressor crosstalk in prediction of the same outcome, the information flux from stressor to stressor also suggests a causal network.

This study achieves the objective by using the Adaptive Lasso regression. Specifically, the Lasso (Least Absolute Shrinkage and Selection Operator) is a linear model estimation method that leverages the sparsity assumption[4]. In common statistical models, consider a linear regression model with a dependent variable Y and hundreds of independent covariates X. The sparsity assumption suggests that only a few number of X's regression coefficients are non-zero[5]. Lasso addresses an issue of identifying importance under the conditions of the high-dimensional variables (a great number of multiplicity) and the sparsity assumption[6, 7]. Theoretical evidences have demonstrated that the Irrepresentable Condition restricts the correlation between the important variables and others[8]. In response to the issue of inconsistency in variable selection encountered by the traditional Lasso, the Adaptive Lasso was proposed[9], which adds data-driven weights to the Lasso penalty and verified by cross-validation. As a result, only the most important covariates can be selected in terms of model selection consistency and parametric estimation asymptotic normality.

Different from the other commonly used algorithm for dimensionality reduction, the Lasso approaches can estimate each feature individually. In many applied scenarios, the variables are *a-priori* grouped and exhibit overlapping foreseeability. On the *a-priori* grouping information, the Group Lasso was firstly introduced in 2006[10], and then the Priority-Lasso[11] was suggested recently that can further rank the candidate stressors as the hierarchical covariates. However, Priority-Lasso provides the priority order on the subjectively selected stressors and some of them can be readily ordered, such an order is not necessarily the causal sequence either.

Cardiovascular disease (CVD) refers to pathologic conditions affecting the heart and blood vessels, responsible for approximately 20.5 million deaths worldwide in 2021[12]. CVD imposes a substantial burden of morbidity and mortality, posing a major threat to human health and well-being[13]. In addition to the well documented SNPs[14], the various environmental contribution includes indoor fuel combustion, second-hand tobacco smoke, ambient air pollution, noise, chemicals, land use and built environment, occupation and climate change[15], the variables' attribution may be overlapped. Among the most commonly reported non-medical stressors of CVD, such as unhealthy diet[16], lack of physical activity[16], tobacco use[17] and harmful use of alcohol[18], it is often challenging to clarify their hierarchical priority in exposure pathways leading to CVD. Although the influence of multiple stressors on cardiovascular health is well-documented[19,8], further analysis is needed to understand how these stressors interacted and contribute to the disease. This study aims to the development of an automated statistical approach for causal path analysis of the multiple stressors of CVD using NHANES datasets, in which an innovative implicational tool called Path-Lasso offers robust support for uncovering the network interaction. The present study will be crucial for optimizing preventive strategies, designing targeted interventions and so on[20], especially for the multiple factors that are crosstalk or prediction overlap when associated with a defined adverse outcome[21].

## 2 Methods
### 2.1 Principles of Path-Lasso

The Path-Lasso is designed to reconstruct the network of candidate variables in the process of predicting specific adverse outcomes based on the Adaptive Lasso (Figure 1). The construction process is as follows: Firstly, the Lasso model is fitted to a set containing all candidate stressors. The objective of this step is to identify the subgroup of variables with the highest priority for predicting CVD. This subgroup explains the greatest amount of variability. Importantly, the result of Lasso model for this first subgroup is consistent with the result obtained when fitting the model to all candidate stressors. Secondly, the model is fitted to a subset that contains all remaining stressors. The objective of this step is to identify the next subgroup of stressors with the second highest priority for prediction. Similarly, the model fitted to this second subgroup variables along produce results that are consistent with those obtained from the remaining variables after the first step. This process is repeated iteratively until no predictive variables remain. The Path-Lasso method stratifies candidate stressors based on their predictive ability using Adaptive Lasso.

The Path-Lasso operates differently from Priority-Lasso when applied to grouped regression screening. The Path-Lasso is a non-prioritizing (?) method that drives by data variation for grouping the candidate stressors in a hierarchal way. When screening out the first layer of variables, the second layer variables are excluded from the model due to the redundancy or collinearity. When screening the second layer variables, the first layer variables are absent and allowing all the second important stressors to be included in the model, thereby, the predictive information of the second layer of variables is entirely hidden under the first layer. The similar correlation is true for the third layer and the second layer. When the pattern continues, each layer encompasses the predictive information of the previous one.

Utilizing lower-level individual variables as outcomes, further regression analysis is conducted using Adaptive Lasso to assess the predictive validity of the higher-level variables to each of the lower-level variables. Thus, the inter-layer associations among the stressors can be constructed. The flow of information can be visualized by moving from the first-tier variables to the outcome through intermediate layers, eventually forming directional paths that create a predictive variable network.

To illustrate the inter-layer causality, we say there is a simple two-layer Path-Lasso network, in which X is defined as a function of all first-layer variables and M is a function of all second-layer variables. In other words, we can call X and M as the latent variables that represent their own inter-layer stressors. It is clear that the coined layer variables of X and M are two stressors of Y either. Formally, a mediation causal model can be constructed among X, M, and Y. Specifically, the Path-Lasso model corresponds to the complete mediation with a path of X→M→Y. Given X→Y is a causal path, we can define the associations among X, M and Y and refer the causal path of X→M→Y, it is because of the two paths have the equivalent causality[22].

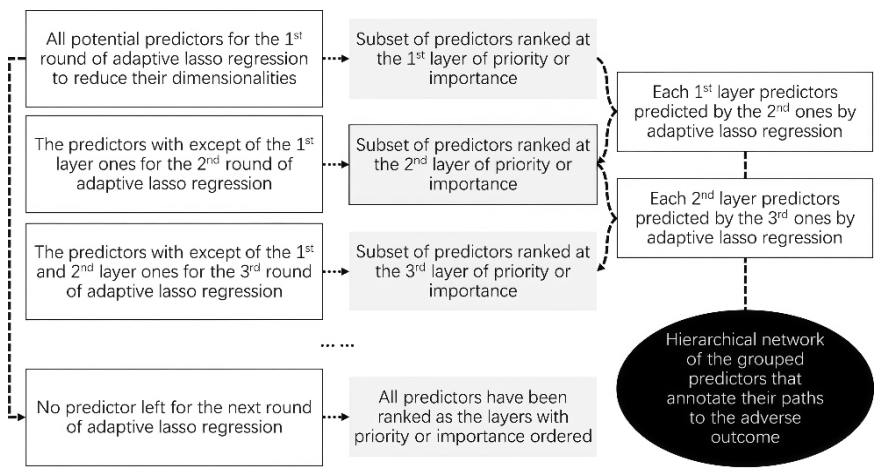

Figure 1 Sparsity hypothesis-based stressor selection regarding of the stressor priority or importance

**2.2 Development of Path-Lasso**

The Adaptive Lasso is an improvement over the Lasso algorithm, aimed at enhancing the predictive accuracy and sparsity of statistical models. Unlike the traditional Lasso method, the Adaptive Lasso introduces adaptive weights into the penalty term, allowing for varying degrees of punishment for each coefficient based on the importance of the features. Let $x_{ij}$ denote the observed value of the j-th variable for the i-th sample (i = 1, …, n), and $y_i$ represent the outcome of the i-th sample. For simplicity, we assume the data is centered, thus the model does not include an intercept term. The Adaptive Lasso method estimates the regression coefficients $\beta_1, …, \beta_p$ by minimizing the following function:

$$\sum_{i=1}^{n}\left(y_i - \sum_{j=1}^{p} x_{ij}\beta_j\right)^2 + \lambda_n \sum_{j=1}^{p} w_j|\beta_j|$$

In the Adaptive Lasso method, the tuning parameter $\lambda_n$ is used to control the degree of sparsity; $w_j = \frac{1}{|\beta_j|^\gamma}, \gamma > 0$, represents the adaptive weights that adjust the penalty size for each variable. These weights can be assigned based on the importance of the features, thereby achieving adaptive feature selection. Typically, the more important features receive the smaller penalties, while the less important features receive the larger penalties. The Oracle property implies that, with high probability, the Adaptive Lasso can identify the correct non-zero coefficients, and the estimates for these coefficients are known a priori. Zou[9] demonstrated that, when an appropriate $\lambda_n$ is chosen, the Adaptive Lasso satisfies the Oracle property.

The coefficient estimates of the Adaptive Lasso satisfy the following conditions, assuming that $\frac{\lambda_n}{\sqrt{n}} \to 0$, $\lambda_n n^{\frac{\gamma-1}{2}} \to \infty$:

(1) Consistency in variable selection: $\lim_n P(A_n^* = A) = 1$, $A = \{j: \beta_j^* \neq 0\}$

(2) Asymptotic normality: $\sqrt{n}(\widehat{\beta_A^{*(n)}} - \beta_A^*) \to N(0, \sigma^2 \times C_{11}^{-1})$

In the Adaptive Lasso, the Least Angle Regression[23] algorithm is employed for computation. LARS is a feature selection method that computationally evaluates the relevance of each feature, progressively calculating the optimal solution through mathematical formulations. LARS initially initializes the weights and the residual vector, then selects the feature vector most correlated with the residual vector, finding a set of weights that minimizes the angle between the new residual vector and this feature vector (i.e., the residual vector is the median vector of the angle between the two vectors). It then recalculates the residual vector. The process continues to find a set of weights that minimizes the angle between the new residual vector and the next most correlated feature vector (i.e., the residual vector is the median vector of the angle between the three vectors), and so on. The iteration process ends when the residual vector is sufficiently small, or all feature vectors have been selected. Although the procedure has computed each feature, in the context of LARS, the results of stressor screening on the same candidate set by the Path-Lasso (Adaptive Lasso) and by the Univariate Logistic Regression may be not consistent.

**2.3 Latent causality between the grouped stressor priorities**

Set a simple scenario that Path-Lasso's regression results are two sets of stressors to correspond the two layers of variables. Set the two-layer variables X and M, which are the functions that represent the first and the second layer variables, respectively. Specifically, the lower (second) layer of variables associated with the outcome in the absence of the upper layer of variables, in other words, they are independent to the outcome in the condition of the upper (first) layer of variables. Given X and M are the causal stressors of the outcome Y. In the equation forms, X and M correspond to the grouped latent layer variable $\eta_{1,i}(1)$ and $\eta_{2,i}(2)$, which are the linear combination of variables in the first and second layer, respectively. When X and M are the validity causes of Y (i.e., X→Y and M→Y), a fully mediated causal path of X→M→Y can be confirmed regarding of the Path-Lasso protocol.

In other words, there must be a complete latent mediation path of X→M→Y if there is casualty of X→Y and M→Y. The above results can be generalized to the Path-Lasso results with more than two hierarchical predictive layers. With the more inter-layer variables be involved into the paths, the approach will bring the more details of intermediate nodes into the X→Y path, i.e., the reduced path will be extended under the causal equivalence[22]. Given the causality of the reduced paths of X→Y, the causality of the extended latent paths of X→M→… …→Y can be inferred.

On the protocol of Adaptive Lasso, each covariate can be either an important stressor or a redundant one. In terms of the neighbored latent variables, the grouped important stressors are always mediated completely by the grouped redundant stressors on the point of their outcome prediction. However, in terms of each stressor, the important stressor is not always mediated by all stressors of the followed redundant groups. The inter-layer stressor connections form a network with path information transit along and to the outcome and offer the profound insights of the stressors' crosstalk.

**2.4 Proof of latent variable causality between the inter-layer grouping covariates**

Assuming that X and M are the validity causes of Y (i.e., X→Y and M→Y), there will be a fully mediated causal path of X→M→Y in the context of the Path-Lasso procedure, which can be proofed as the follows:

X is the direct cause of Y, $P(X,Y) = P(X)P(Y|X)$;

Under the condition of M, X→Y is also causality true, then $P(X,Y) = P(X)P(Y|X,M)$

Then, we can get $P(Y|X) = P(Y|X,M)$     **(1)**

M is the cause of Y, $P(M,Y) = P(M)P(Y|M)$, but it holds only in the absence of X. Under the condition of X, M and Y are independent,

that is $P(M,Y|X) = P(M|X)P(Y|X)$     **(2)**

or equivalently to say:

$P(M|Y,X) = P(M|X)$     **(3-1)**  or

$P(Y|X,M) = P(Y|X)$     **(3-2)**

On the definition of three variable conditional probability:

$P(M,Y,X) = P(M,Y|X)P(X)$     **(4)**

Based on the above facts, the causal correlation between X and M can be deduced step-by-step.

From (4) and (2), we can get (5)

$$\frac{P(M,Y,X)}{P(X)} = P(M,Y|X) = P(M|X)P(Y|X)$$

$$\Rightarrow P(X)P(M|X) = \frac{P(M,Y,X)}{P(Y|X)} \quad (5)$$

Substituting (3-2) into (5) can yields (6):

$$P(X)P(M|X) = \frac{P(Y,X,M)}{P(Y|X,M)} = P(X,M)$$

$$\Rightarrow P(X,M) = P(X)P(M|X) \quad (6)$$

Then the causality of X→M can be finally proofed.

**2.5 NHANES Datasets and Study Subjects**

The concept of Path-Lasso path analysis is tested using the 2015-2018 cycles of the NHANES. NHANES is ongoing population-based survey, conducted by the U.S. Centers for Disease Control and Prevention. NHANES recruit a representative sample of the civilian U.S. population, aiming to assess the environmental chemical exposures and monitor health and nutritional status. All participants included in the NHANES cycles completed the informed consent process. The NCHS Ethics Review Board at the CDC approved the data collection for NHANES. As the present analysis uses publicly de-identified data files, the Human Research Protection Program at Michigan State University deemed the present analyses to not involve human subjects. We utilized the NHANES data excluding the individuals under 20 years old and those with missing data on key variables. The final analytical sample comprised 2,936 adults, as illustrated in Figure 2 (A) depicting the selection process.

**2.6 Study Variables**

The adverse outcome is CVD, which was assessed through the self-reports from participants. The questionnaire inquired about whether a doctor had diagnosed them with congestive heart failure, coronary heart disease, angina, heart attack, or stroke. An affirmative response to any of these conditions indicated a history of CVD. Variables of interest extracted from NHANES included demographic characteristics (age, gender, race, education level, marital status), behavioral characteristics (recreational activities, smoking, alcohol consumption), clinical information (CVD, diabetes, body mass index (BMI), waist circumference), and laboratory measures (blood and urine cadmium, urine creatinine, high sensitivity C-reactive protein).

**2.7 Statistical Analysis**

Balance tests, which included t-tests or rank-sum tests for quantitative variables and chi-square tests or rank-sum tests for categorical variables, were conducted to assess the characteristics of both included and excluded study participants. Continuous variables were expressed as median and categorical variables as percentage. Univariate Logistic regression initially screened CVD-associated factors, followed by Path-Lasso using Adaptive Lasso regression to construct covariate networks. Firstly, all covariates are grouped into one set. Secondly, the importance of each covariate is evaluated by Path-Lasso, it is a stepwise Lasso regression to address the corresponding covariate sparse matrices, and these matrices have the different importance or priority to predict CVD. Finally, the predictive variables will be stratified into layers according to the sparse matrix importance. The first layer of covariates associated with the risk of CVD with the most importance. Meanwhile the second layer of covariates associated with CVD in the absence of the first layer of covariates, in other words, they are independent to CVD in the condition of the first layer of variables. The associations among the second and the third layers' variables are similar to the first and the second ones. Step by step, the associations are continuous in between the neighboring layers with the less and less importance. For each variable in the lower layer that may be or not be predicted by the upper layer variables, therefore, the additional associations were verified by Adaptive Lasso. Finally, the associational network was formed with variables as nodes and associations as edges, indicating paths from covariates to CVD outcomes via some intermediary variables. The path causality can be implied on the upper mention proof.

The variables in the unidirectional paths can be interpreted by generation or reflection of prediction between the neighboring variables, which will finally forward to the CVD risk. The above statistical analyses were accomplished using R 4.3.1 and JMP Pro 16.

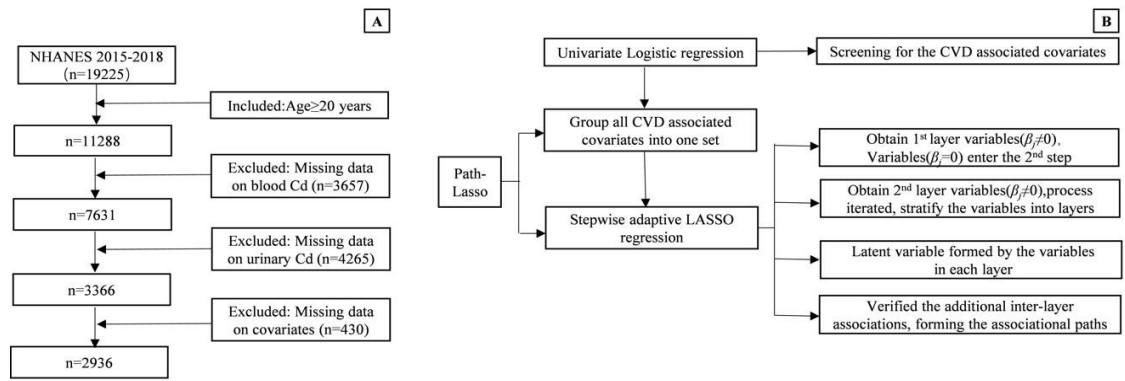

**Figure 2** Flowcharts of participant screening and enrollment (**A**) and the Path-Lasso application for path analysis of the CVD stressors (**B**)

## 3 Results

### 3.1 Participant Baseline and Candidate Stressors of CVD Morbidity

A total of 2,936 eligible subjects were included in this study (Figure 2 A). The balance analysis for the inclusion and exclusion was performed (Table S1). The weighted prevalence of CVD in the overall U.S. population was 7.9%. Other baseline characteristics of study subjects are described in Table S2.

The preliminary exploration of the relationship between the factors and CVD was conducted by the univariate logistic regression. The risk of CVD morbidity was associated with 11 covariates, which are the cadmium in blood (Blood Cd) and urine (Urinary Cd), aging (Age), races (Non-Hispanic White, Non-Hispanic Black, Mexican American and Other Hispanic for short NHW, NHB, MA and OH, respectively), marital status, recreational activity, smoking status, drinking status, diabetes, waist circumference (Waist), hypersensitive C-reactive protein (HSCRP) ($P < 0.05$) (Table S3).

### 3.2 Multiple Covariates Associated with CVD Hierarchically

Path-Lasso analyses were performed on the subjects that stratified by smoking status, i.e., former smokers *vs.* never smokers and current smokers *vs.* never smokers, respectively. It is regarding the fact that smoking can associate CVD with much more importance and its different status associated with CVD via the different intermediary covariates. The summary results are presented in Table 1 (with details in Table S4-Table S11), in which the results showed that Waist and Age were involved in the former smoker model and the Current smoking and Age entered the current smoker model, respectively. Amogh the investigated multiple variables, a few of them are with the highest priority or most importance to predict CVD in the related models. The highest priority variables can be grouped and taken as the latent variables, which are the linear combination of Waist and Age in the former smoker model, and Current smoking and Age in the current smoker model.

Because the impaction of these grouped first layer variables to CVD has been completely transited by the rest variables, the second step regression on the rest stressors will remove the redundant information. Then the Adaptive Lasso was performed on the remaining variables to screen out the second layer variables. Step by step, the third layer of variables were identified, and finally there is no variable that can be observed as the fourth layer variable. All in one, three hierarchical layers of variables have been addressed for the two models.

**Table 1.** Adaptive Lasso Identified Hierarchical Covariate Associated with CVD

| Sample | Stepwise Adaptive Lasso | |
|---|---|---|
| Former smoker *vs.* Never smoker | First layer | Waist, Age |
| | Second layer | Ln (Urinary Cd), Former smoking, Race, Marital status, Recreation activity |
| | Third layer | Ln (Blood Cd) |
| Current smoker *vs.* Never smoker | First layer | Current smoking, Age |
| | Second layer | Ln (Urinary Cd), Waist, Race, Recreation activity |
| | Third layer | Ln (Blood Cd), Marital status, Diabetes |

**3.3 Upper Layer Covariates Associated CVD via Lower Layer Ones**

Based on the hierarchical layered variables (Table 1), the neighboring layer variables' association was explored by Adaptive Lasso. The regression was separately done on the two subjects' smoking statuses. The operation is that all variables in the upper layer were regressed to each variable in its neighboring lower layer. The calculation factually assessed the ability of the variables in the upper layer to predict the variables in the next lower layer.

The results are presented in Table 2 (with details in Table S12 and Table S13) for the former smokers, and in Table 3 (with details in Table S14 and Table S15) for the current smokers. The inter-layer variable association confirmed the pairwise partial transition of the CVD risk. To the former smokers, Age and Waist are the biggest stressors of CVD, the risk prediction carried by the two variables can transit via different mediation paths of the CVD risk. For example, from obesity (Waist) via less excises (Recreation activity) to CVD. It is interesting that aging (Age) can predict the former smoking risk to CVD, which indict that the subjects' historical records cannot be omitted in the risk assessment. For the current smoking subjects, it is clear that smoking moved up to the first importance among the stressors. Current smoking and aging are the two biggest and independent stressors to CVD morbidity, the former have associated $X_{RA}^V$, $X_u^{Cd}$ and $X_{race}^{NHW}$.

On the Path-Lasso principles, the latent variable of the upper layer stressors ordered higher priority than the latent variable of the lower layer stressors, which completely linked to the risk of CVD in causality in step-by-step. In the causal context, each upper layer stressor can at least partially predict some of the lower layer stressors in causality. All the observed linkages are on the stepwise ways of CVD prediction. Regarding the association of the first latent variable with the outcome, the intermediary latent variables are transitive closure and can be reduced to some extent, which means only aging, obesity and/or smoking habit are the major stressors of CVD (only the last two are directly preventable).

**Table 2.** Analysis of the interlayer variables' connection (**Former smoker**)

| Outcome | Inclusion and exclusions |
|---|---|
| **Second layer** | **First layer of variables: Waist, Age** |
| Ln (Urinary Cd) | Both Waist and Age were included |
| Former smoker | Both Waist and Age were included |
| Race | Reference：Other/multiracial <br> ● In Mexican American, Other Hispanic and Non-Hispanic Black, Waist is included, while Age is not included <br> ● In Non-Hispanic White, both Waist and Age were included |
| Marital status | Reference：Married <br> In Unmarried and Dissolved marriage, Age is included, Waist is not included |
| Recreation activity | Reference：No or lower <br> ● In Vigorous level, both Waist and Age were included <br> ● In Moderate level, Waist and Age were not included |
| **Third layer** | **Second layer of variables: Ln (Urinary Cd), Former smoker, Race, Marital status, Recreation activity** |
| Ln (Blood Cd) | ● Ln (Urinary Cd), Former smoking, Race [Mexican American, Other Hispanic, Non-Hispanic White, with Other/multiracial as reference] were included <br> ● Race [Non-Hispanic Black, with Other/multiracial as reference], Marital status and Recreation activity were not included |

**Table 3.** Analysis of the interlayer variables' connection (**Current smoker**)

| Outcome | Inclusion and exclusions |
|---|---|
| **Second layer** | **First layer of variables: Current smoker, Age** |
| Ln (Urinary Cd) | Both Current smoking and Age were included |
| Waist | Age was included and Current smoking was excluded |
| Race | Reference：Other/multiracial <br> ● In Non-Hispanic White, Current smoking and Age were included <br> ● In Non-Hispanic Black, only Current smoking is included and Age is excluded <br> ● Mexican American and Other Hispanic, Current smoking and Age are not included |
| Recreation activity | Reference：No or lower |
| | - In Vigorous level, Current smoking and Age are included<br>- In Moderate level, Current smoking and Age are not included |
| **Third layer** | **Second layer of variables: Ln (Urinary Cd), Waist, Race, Recreation activity** |
| Ln (Blood Cd) | Ln (Urinary Cd) and Race were included, and Waist and Recreation activity were excluded |
| Marital status | Reference：Married<br>- In Unmarried, Ln (Urinary Cd), Waist, Race [Non-Hispanic Black-Other/multiracial] are included, Race [Mexican American, Other Hispanic, Non-Hispanic White, with Other/multiracial as reference], Recreation_ activity is not included<br>- In dissolved marriage, Ln (Urinary Cd), Race [Other Hispanic, Non-Hispanic White, Non-Hispanic Black, with Other/multiracial] were included, and Waist, Race [Mexican American, with Other/ multiracial] and Recreation activity were not included |
| Diabetes | Waist, Race [Non-Hispanic White-Other/multiracial], Recreation activity[Vigorous-No or lower] were included, and Ln (Urinary Cd), Race[Mexican American, Other Hispanic, Non-Hispanic Black, with Other/ multiracial] were excluded |

**3.4 Covariate Paths to Predict CVD Morbidity**

The hierarchal layer variables' estimators (i.e., the latent variables of $\hat{\eta}_{1,i}(1)$, $\hat{\eta}_{2,i}(2)$ and $\hat{\eta}_{3,i}(3)$) have sequenced the grouped variables' priorities. Meanwhile, the covariate inter-layer connections have formed the network with all the stepwise paths from the layer variables to CVD morbidity, in which the variables in layers are network nodes and their inter-layer correlations are network edges. Because all covariates (nodes) are validated stressors of the CVD risk, their connectional paths (i.e., sequenced edges) to the outcome are completely mediated by some intermediary covariates. The network for the former smokers is shown in Figure 3, and the network for current smokers is presented in Figure 4. The corresponding β values indicate the regression coefficients of the stressors when they predict CVD. Among all covariates, Recreation activity ($X_{RA}^V$ in Figure 3 and 4, and notes in Table S4 and S5) is the only one that is negatively associated with the risk. With the networks and paths, it is easy to see how the variables in the lower layer are predicted by the upper layer variables in the CVD context. For example, Marital status involves the path of Age to CVD but not included in the path of Waist to CVD, it implies that aging plus to bad marriage can increase the CVD risk, but obesity cannot.

Beyond the assessment of the stressors' hierarchical importance, the major exposure paths can be visualized either. For the former smokers, we may say that four major exposure paths of Recreation activity, Ln (Blood Cd), Marital status and Race of Non-Hispanic Black have impacted the CVD risk; for the current smokers, the major paths are diabetes, Ln (Blood Cd) and Marital status. More or less, these major paths can be modified by the other factors behind them.

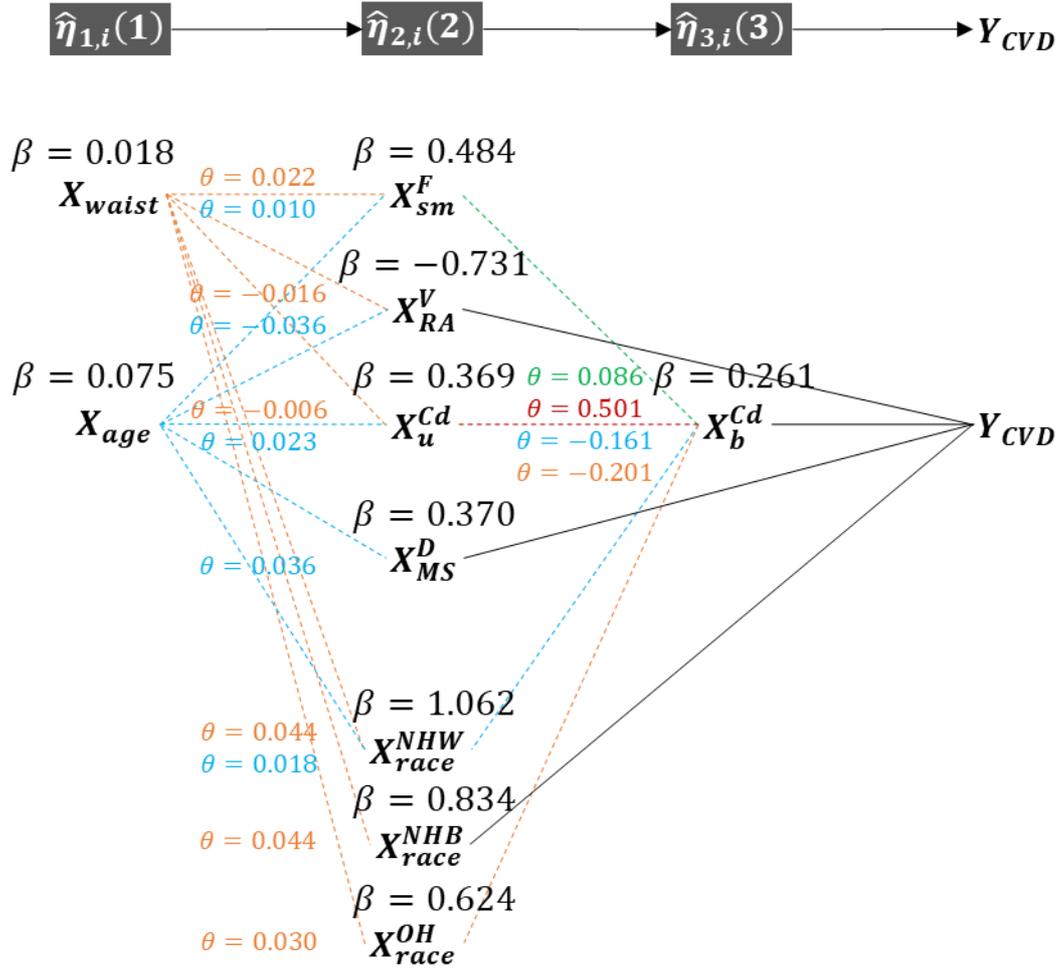

**Figure 3** Latent layer variable causality and hierarchal stressor network with predictive paths of CVD in the Former smokers

$\hat{\eta}_{1,i}(1)$, $\hat{\eta}_{2,i}(2)$ and $\hat{\eta}_{3,i}(3)$ are the three latent causal layer variables' estimators; $X^F_{sm}$: Former smoking, with Never smoking as reference; $X^V_{RA}$: Recreation activity (Vigorous, with No or lower as reference); $X^{Cd}_u$: Ln (Urinary Cd); $X^D_{MS}$: Marital status (Dissolved marriage, with Married as reference); $X^{NHW}_{race}$: Race(Non-Hispanic White, with Other/multiracial as reference); $X^{NHB}_{race}$: Race (Non-Hispanic Black, with Other/multiracial as reference); $X^{OH}_{race}$: Race (Other Hispanic, with Other/multiracial as reference); $X^{Cd}_b$: Ln (Blood Cd); $Y_{CVD}$:Heart disease; $\beta$: Correlation coefficient of $X$ associated with $Y_{CVD}$; $\theta$: Correlation coefficient of the upper layer $X$ associated with the neighbored lower layer $X$ indicated by the dotted line with the same color. The factor X can associate with $Y_{CVD}$ both directly (with correlation coefficient $\beta$) and via one or more mediatory X (with correlation coefficient $\theta$) is distal stressor; while others can associate with $Y_{CVD}$ directly only is proximal stressor. The proximal stressors (i.e., $X^V_{RA}$, $X^D_{MS}$, $^{HB}_{race}$ and $X^{Cd}_b$) are closer to the outcome of CVD than the distal ones.

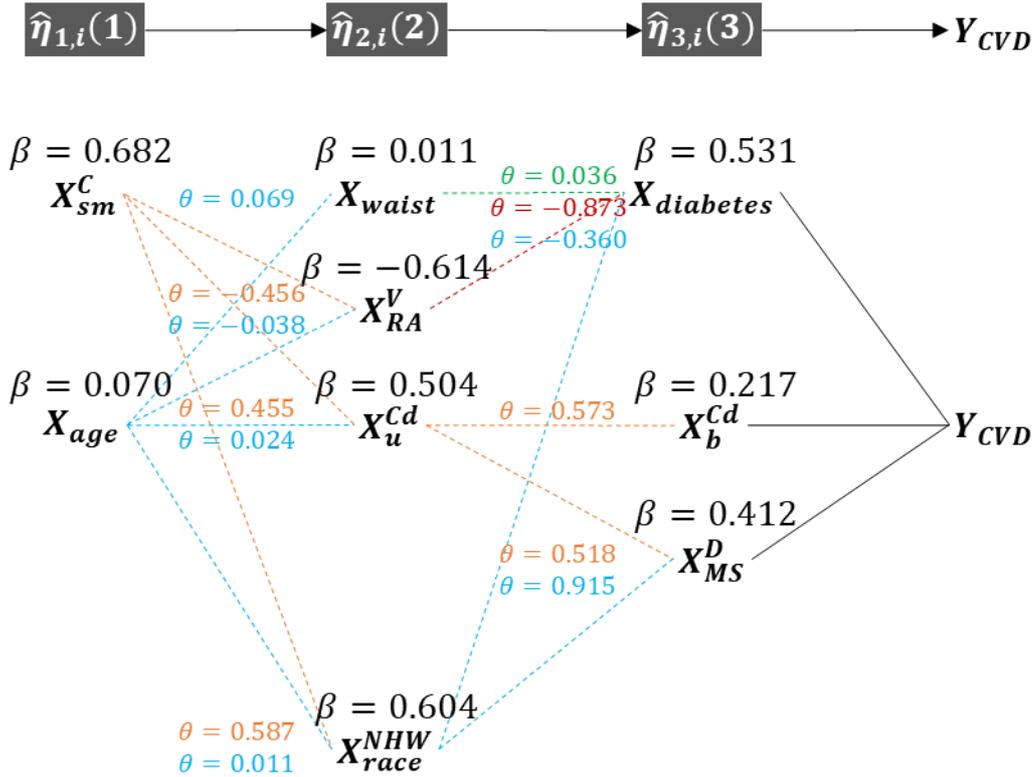

**Figure 4** Latent layer variable causality and hierarchal stressor network with predictive paths of CVD in the Current smokers

$\hat{\eta}_{1,i}(1)$, $\hat{\eta}_{2,i}(2)$ and $\hat{\eta}_{3,i}(3)$ are the three latent causal layer variables' estimators; $X_{sm}^{C}$: Former smoking, with Never smoking as reference; $X_{RA}^{V}$: Recreation activity; $X_{u}^{Cd}$: Ln (Urinary Cd); $X_{race}^{NHW}$: Race (Non-Hispanic White, with Other/multiracial as reference); $X_{b}^{Cd}$: Ln (Blood Cd); $X_{MS}^{D}$: Marital status (Dissolved marriage, with Married as reference); $Y_{CVD}$: Heart disease; $\beta$: Correlation coefficient of $X$ associated with $Y_{CVD}$; $\theta$: Correlation coefficient of the upper layer $X$ associated with the neighbored lower layer $X$. The factor X can associate with $Y_{CVD}$ both directly (with correlation coefficient $\beta$) and via one or more mediatory X (with correlation coefficient $\theta$) is distal stressor; while others can associate with $Y_{CVD}$ directly only is proximal stressor. The proximal stressors (i.e., $X_{diabetes}$, $X_{MS}^{D}$ and $X_{b}^{Cd}$) are closer to the outcome of CVD than the distal ones.

## 4 Discussion

In epidemiological studies, the crosstalk between multiple stressors along disease prediction pathways have posted challenges for quantification and interpretability, especially when certain covariates overlapping information. Many studies have only focused primarily on covariate selection by reducing redundancy[24, 25]. However, the overlapping features of covariates can be valuable for tracing their influence on the outcome. To address this issue, we developed the Path-Lasso approach, which hierarchically groups covariates in a data driven manner The method first screens all potential stressors and cognizes then into groups based on their priority to predict the outcome. The sequentially ordered stressors can be linked step-by-step to the outcome. We applied this approach to the NHANES dataset to trace which factors and how they affect cardiovascular disease (CVD) risk. The candidate covariates involved in the present work are lifestyle factors, environmental exposures, and clinical indicators, our results provide an in-depth interpretability of how these factors impact CVD risk.

Adaptive Lasso is useful for variable selection by eliminating redundancy that includes overlapping and unpredictable variables[26, 27]. Overlapping covariates are those that less important due to their collinearity with the high priority covaraites[28]. It implies that when the different factors are applied to feature the same outcome, the variable redundancy comes from their complete or partial mediational variation to the targeted outcome[29]. On this fact, the Path-Lasso groups stressors based on their predictive importance, the connected neighboring layers form a hierarchical network with covariate paths of the outcome. Different from the Group Lasso[30] and Priority Lasso[31] that regress the candidate variables on their a-priori categories, Path-Lasso treats all candidates equally and constructs the hierarchical layers automatically. Lower-layer variables are associated with the outcome only when upper-layer variables are absent from the same regression. The Lasso method is also over other machine learning techniques primarily due to its ability to automatically screen variables under the assumption of sparsity and produce interpretable results, which closely aligns with our objective of identifying the factors of CVD risk. Compared to black-box models such as random forests and XGBoost, Lasso offers greater interpretability and avoids the issues of potential overfitting and outlier sensitivity that these models may encounter. Additionally, when we facilitated its unique dimensionality reduction to construct the stressors' hierarchical network and paths, it effectively addresses the issue of the covariates' priority orders when the outcome is predicted.

The latent layer variables have been proofed to exhibit causal relationships, with the first layer latent serving as a progenitor and subsequent layers as the posteriors. The lower-layer variables mediate the association between the neighboring upper layer variables and the outcome (see the X→M→Y model in the section 2.3), which make the priority orders may be causally important that arisen from a cross-sectional dataset. However, the hypotheses of both X and M are the cause of Y is hard to meet, it is because all covariates in each layer are required to be the causality of outcome. For example, we may accept that aging and smoking (covariates in the 1$^{st}$ latent layer variable) are the causes of the CVD risk, but race factor (covariates in the 2$^{nd}$ latent layer variable) may not be true (Figure 4).

The Path-Lasso can be used to annotate crosstalk of the multiple stressor exposure of an adverse outcome. Univariate analysis can usually identify a broad of heterogeneous stressors like age, race, lifestyle, biomarkers with existing outcome[32-43]. This work profiles the hierarchical networks of the stressors. In the CVD case, when all the 11 stressors were modeled together, only Age, Waist and current smoking emerged as the most important stressors. It is interesting that when subjects involved into the model with the smoking status that stratified as current smokers, former smokers and never smokers (as the reference), only the current smoking can predict CVD, suggesting differences between current and former smokers[44]. The present work showed that current smoking and former smoking may not only be characterized difference in smoking intensity, but also in paths of causing CVD morbidity.

The Path-Lasso networks may be used to interpret the epidemiological mechanism (or stressor paths to adverse outcome) in two ways: 1) Tracing the paths from the distal or highest priority stressors to the outcome, showing the step-by-step associations mediated by some intermediary stressors. 2) Identifying major paths from the proximal stressors, which at least partially have been modified by the upstream stressors. The insights would be useful for disease prevention design.

In the present results, waist circumference and age are the two most important CVD stressors in the former smokers. Staring at Waist and Age, there are 11 different paths associated with the CVD. For examples, the "Age-former smoking-blood cadmium-CVD" path implies blood cadmium as a risk factor modified by aging and past smoking exposure. It is clear that cigarette exposure is one of the major sources of cadmium[45, 46], and the old people may have exposed smoking longer than the young people. It is interesting to say that Recreation activity (in vigorous level), Marital status (in dissolved marriage) and cadmium exposure may be three independent risk factors. Vigorous activity is the only protective factor, which can mitigate the risk from both obesity and aging. However, aging but obesity has modified the risk of poor marital status on CVD, suggesting of the psychological support are required for the aging people[47]. We may say that the prediction of urinary cadmium to the blood cadmium reveals their reflection but generation tie. Cadmium in blood can correspond to the CVD outcome more rational than cadmium in urinary matrix. With more details, we can say the blood cadmium has been modified by the more covariates; especially the former smoking habit has tied with blood cadmium but urinary cadmium (Figure 3); meanwhile the current smoking can directly associate with urinary and indirectly with blood cadmium (Figure 4). In addition, some paths are significant only in race special patterns, especially obesity has modified more race status than aging. To the former smokers, Blood cadmium, Marital status (in dissolved marriage), Race (in NHB) and Recreation activity (in vigorous level) are the most closed factors to predict the CVD risk, while the others are more or less distal ones to predict CVD (Figure 3).

To the current smokers, smoking and aging are two most important factors to predict the CVD. The path of "Age-Urinary cadmium-Blood cadmium-CVD" occurred again in these subjects. Specifically, diabetes plays as an additional risk of CVD. For the current smokers, diabetes, Blood cadmium and Marital status (in dissolved marriage) are three proximal stressors of CVD among the investigated covariates. It seems to be tricky that diabetes plays as the risk of CVD in the current smokers[48]. When comparison to former smoking, current smoking may has counteracted the benefit of Recreation activity and open the related path [49, 50](Figure 3 and 4). Regardless of the smoking status, Recreation activity is benefit but Blood cadmium is hazardous to prevent the CVD, which is more direct to associate with CVD than the others. In addition，HSCRP has be observed to associate with CVD but is absent in the Path-Lasso models, which may imply this factor is either the result of CVD or impact CVD via the other way[51, 52]. All interpretation is tentative and needs to be verified in the further study.

In summary, the research developed a programmed protocol, called the Path-Lasso, to deconstruct the intricate paths of connecting of various stressors to predict adverse outcomes. We found the automated method is able to delineate the network of predictive covariates with the redundant risk information transitable toward CVD, offering profound insights into the stressors' impact paths. The Path-Lasso is efficient to rank the high dimensionality variables' priorities in the parsimony hypothesis on their predictive power, while also providing the latent causal inference on the grouped stressor priorities regardless of the unmeasured confounders, ensuring the validity of quantitative partial causality claim with regarding the covariates' crosstalk. The results may help the setting of preventable checkpoints in disease prevention powerful.

**5 Strengths, Limitation and Outlook**

The present work has offered Path-Lasso as a novelty approach to explore the epidemiological exposure pathways on a defined outcome. When an outcome is associated with the multiple covariates, the high dimensional variable subsets can be used to construct a hierarchical network, where the covariates in each layer indicate the different importance for predicting the outcome. The stressors are identified and stratified using the Adaptive Lasso with Oracle Properties which helps validate the robustness of the Path-Lasso results. The Path-Lasso is a general approach, and therefore may be used to explore the molecular paths of the multiple OMICs data for in-depth understanding of both observation and laboratory studies. What's more, the Path-Lasso approach takes into account not only single factors, but also the interactive effects among multiple factors, which contribute to a more comprehensive understanding of the risks of diseases. However, there are some limitations in this study. First, causal associations based on individual variables require further validation. Second, the interpretability of the pathways needs to be confirmed through the validation results, and future experimental studies will be needed. Upon further validation, the Path-Lasso could significantly advance and support the related research.

**Funding**

None.

**Conflict of interest**

The authors have no conflicts of interest

**Data Sharing**

The data underlying this article are available in github, at https://github.com/1818036/Data-Code/tree/main. The datasets were derived from sources in the public domain: NHANES, https://www.cdc.gov/nchs/nhanes/index.htm.

**Author Contributions**

**Heqing Shen:** Conceptualization, Investigation, Methodology, Project administration, Resources, Supervision, Writing – original draft, Writing – review and editing. **Longjian Liu**: Conceptualization, Methodology, Project administration, Supervision, Writing – review and editing. **Jiangling Liu**: Conceptualization, Data curation, Formal analysis, Investigation, Methodology, Software, Visualization, Writing – original draft, Validation. **Ya Liu:** Conceptualization, Data curation, Formal analysis, Investigation, Methodology, Project administration, Software, Visualization, Writing – original draft, Validation. **Banyun Zheng:** Investigation, Methodology, Resources, Software, Visualization, Validation.

**Ethics approval and informed consent**

This study utilized publicly available data from the National Center for Health Statistics, which does not require institutional review board (IRB) or ethics committee approval. The data are accessible at https://www.cdc.gov/nchs/nhanes/index.htm. As the data are de-identified and already in the public domain, informed consent was not applicable.

**Supporting Information for**

**A Protocol to Exposure Path Analysis for Multiple Stressor's Associated with CardiovascularDisease Risk: A Novel Approach Using NHANES Data**

Jiangling Liu[a,1], Ya Liu[a,1], Banyun Zheng[a,1], Longjian Liu[b,2], Heqing Shen[a,2]

[1]Jiangling Liu, Ya Liu and Banyun Zheng contributed equally to this work.
[2]To whom correspondence may be addressed. Email: hqshen@xmu.edu.cn.

Author affiliations: [a] State Key Laboratory of Vaccines for Infectious Diseases, Xiang An Biomedicine Laboratory & State Key Laboratory of Molecular Vaccinology and Molecular Diagnostics, School of Public Health, Xiamen University, Xiamen, 361102, China; [b] Department of Epidemiology and Biostatistics, Dornsife School of Public Health, Drexel University, Philadelphia, PA, 19104, USA.

Corresponding authors:
Heqing Shen, School of Public Health, Xiamen University, 4221-117 Xiang'an South Road, Xiamen, 361102, China; Email: hqshen@xmu.edu.cn.
ORCID of the author：https://orcid.org/0000-0002-6544-4919.
Longjian Liu: Drexel University, Longjian.Liu@Drexel.edu

Co-First-author:
Jiangling Liu: Xiamen University, liujiangling@stu.xmu.edu.cn
Ya Liu: Xiamen University, 32620221150850@stu.xmu.edu.cn
Banyun Zheng: Xiamen University, zhengbanyun@stu.xmu.edu.cn

**This word file includes:**

    Supporting text
    Tables S1 to S15

**Supporting Information Text**

**Subhead.** The Supporting Information appendix contains comprehensive tables that present the outcomes of the study's various analytical procedures. These include the equilibrium test for the selection of study participants, descriptive statistics of the study population's general characteristics, the results of univariate logistic regression analysis, and the step-by-step outcomes of the path-lasso analysis conducted on the two distinct samples. This appendix is provided to illuminate the analytical process employed throughout the study, ensuring transparency and a thorough understanding of the methodological framework that underpins the research findings.

# Supplementary Notes
## CONTENTS

**Table S1.** Balance Assessment of the Inclusion and Exclusion Criteria

| Characteristics | Exclude (N=8352) | include (N=2936) | Total (N=11288) | P-value |
|---|---|---|---|---|
| **Age, years** | | | | 0.003 |
| 20-39 | 2715 (32.5%) | 925 (31.5%) | 3640 (32.2%) | |
| 40-59 | 2637 (31.6%) | 960 (32.7%) | 3597 (31.9%) | |
| 60-79 | 2366 (28.3%) | 882 (30.0%) | 3248 (28.8%) | |
| 80+ | 634 (7.6%) | 169 (5.8%) | 803 (7.1%) | |
| **Sex** | | | | 0.021 |
| Female | 4374 (52.4%) | 1465 (49.9%) | 5839 (51.7%) | |
| Male | 3978 (47.6%) | 1471 (50.1%) | 5449 (48.3%) | |
| **Race** | | | | 0.180 |
| Mexican American | 1265 (15.1%) | 465 (15.8%) | 1730 (15.3%) | |
| Other Hispanic | 944 (11.3%) | 341 (11.6%) | 1285 (11.4%) | |
| Non-Hispanic White | 2776 (33.2%) | 1022 (34.8%) | 3798 (33.6%) | |
| Non-Hispanic Black | 1872 (22.4%) | 624 (21.3%) | 2496 (22.1%) | |
| Other/multiracial | 1495 (17.9%) | 484 (16.5%) | 1979 (17.5%) | |
| **Education attainment** | | | | 0.206 |
| N-Miss | 18 | 0 | 18 | |
| ≤9th | 886 (10.6%) | 281 (9.6%) | 1167 (10.4%) | |
| 9-11th | 978 (11.7%) | 336 (11.4%) | 1314 (11.7%) | |
| High School | 1854 (22.2%) | 707 (24.1%) | 2561 (22.7%) | |
| College or AA degree | 2577 (30.9%) | 893 (30.4%) | 3470 (30.8%) | |
| College Graduate+ | 2039 (24.5%) | 719 (24.5%) | 2758 (24.5%) | |
| **Marital status** | | | | 0.192 |
| N-Miss | 9 | 0 | 9 | |
| Married | 4913 (58.9%) | 1780 (60.6%) | 6693 (59.3%) | |
| Dissolved marriage | 1905 (22.8%) | 627 (21.4%) | 2532 (22.4%) | |
| Unmarried | 1525 (18.3%) | 529 (18.0%) | 2054 (18.2%) | |
| **BMI group** | | | | 0.870 |
| N-Miss | 707 | 0 | 707 | |
| <18.5 | 125 (1.6%) | 44 (1.5%) | 169 (1.6%) | |
| 18.5-25 | 1973 (25.8%) | 742 (25.3%) | 2715 (25.7%) | |
| 25-30 | 2455 (32.1%) | 943 (32.1%) | 3398 (32.1%) | |
| ≥30 | 3092 (40.4%) | 1207 (41.1%) | 4299 (40.6%) | |
| **Recreational activity** | | | | 0.169 |
| N-Miss | 4 | 0 | 4 | |
| Moderate | 1908 (22.9%) | 701 (23.9%) | 2609 (23.1%) | |
| No or lower | 4459 (53.4%) | 1509 (51.4%) | 5968 (52.9%) | |
| Vigorous | 1981 (23.7%) | 726 (24.7%) | 2707 (24.0%) | |
| **Drinking status** | | | | 0.298 |
| N-Miss | 1449 | 0 | 1449 | |
| Light drinker | 2892 (41.9%) | 1244 (42.4%) | 4136 (42.0%) | |
| Heavy drinker | 794 (11.5%) | 370 (12.6%) | 1164 (11.8%) | |
| Moderate drinker | 481 (7.0%) | 207 (7.1%) | 688 (7.0%) | |
| Non-drinker | 2736 (39.6%) | 1115 (38.0%) | 3851 (39.1%) | |
| **Smoking status** | | | | **0.056** |
| N-Miss | 11 | 0 | 11 | |
| Current smoker | 1524 (18.3%) | 550 (18.7%) | 2074 (18.4%) | |
| Former smoker | 1917 (23.0%) | 731 (24.9%) | 2648 (23.5%) | |
| Never smoker | 4900 (58.7%) | 1655 (56.4%) | 6555 (58.1%) | |
| **Diabetes** | | | | **0.029** |
| N-Miss | 1069 | 0 | 1069 | |
| NO | 6222 (85.4%) | 2557 (87.1%) | 8779 (85.9%) | |
| YES | 1061 (14.6%) | 379 (12.9%) | 1440 (14.1%) | |
| **Blood Cadmium** | | | | **0.797** |
| N-Miss | 3657 | 0 | 3657 | |
| Mean (SD) | 0.486 (0.564) | 0.482 (0.565) | 0.485 (0.564) | |
| Range | 0.070 - 13.030 | 0.070 - 7.500 | 0.070 - 13.030 | |
| **Urinary Cadmium** | | | | **<0.001** |
| N-Miss | 7786 | 0 | 7786 | |
| Mean (SD) | 0.416 (0.530) | 0.339 (0.391) | 0.351 (0.417) | |
| Range | 0.025 - 7.581 | 0.025 - 4.247 | 0.025 - 7.581 | |
| **Age** | | | | **0.810** |
| Mean (SD) | 50.531 (18.014) | 50.439 (17.233) | 50.507 (17.814) | |
| Range | 20.000 - 80.000 | 20.000 - 80.000 | 20.000 - 80.000 | |
| **Waist** | | | | **0.367** |
| N-Miss | 1235 | 0 | 1235 | |
| Mean (SD) | 100.386 (16.869) | 100.719 (16.657) | 100.483 (16.807) | |
| Range | 57.900 - 171.600 | 58.700 - 164.900 | 57.900 - 171.600 | |
| **HSCRP** | | | | **0.021** |
| N-Miss | 1246 | 0 | 1246 | |
| Mean (SD) | 4.392 (8.793) | 3.973 (6.748) | 4.269 (8.250) | |
| Range | 0.080 - 188.500 | 0.080 - 94.100 | 0.080 - 188.500 | |
| **CVD** | | | | **0.020** |
| Yes | 7336 (87.8%) | 2626 (89.4%) | 9962 (88.3%) | |
| NO | 1016 (12.2%) | 310 (10.6%) | 1326 (11.7%) | |

**Note:** N-Miss: Missed number of participants; Mean (SD): Mean (standard deviation); HSCRP: Hypersensitive C-reactive protein; CVD: Cardiovascular disease. In the data cleaning phase, we compared excluded and included records. The comparison revealed no significant differences across most variables between the two datasets, suggesting that our sample remains well-balanced overall.

**Table S2.** Baseline Characteristics of Study Participants

| Characteristics | Participant (N = 2936[1]) | Characteristics | Participant (N = 2936[1]) |
|---|---|---|---|
| Blood Cd | | Recreation activity | |
| Q1 | 651 (28%) | No or lower | 1,509 (45%) |
| Q2 | 614 (23%) | Moderate | 701 (26%) |
| Q3 | 785 (25%) | Vigorous | 726 (29%) |
| Q4 | 886 (25%) | BMI group | |
| Urinary Cd | | <18.5 | 44 (1.4%) |
| Q1 | 572 (25%) | 18.5-25 | 742 (25%) |
| Q2 | 715 (25%) | 25-30 | 943 (31%) |
| Q3 | 708 (25%) | ≥30 | 1,207 (43%) |
| Q4 | 941 (25%) | Smoking status | |
| Age group | | Never smoker | 1,655 (56%) |
| 20-39 years | 925 (36%) | Former smoker | 731 (27%) |
| 40-59 years | 960 (36%) | Current smoker | 550 (17%) |
| 60-79 years | 882 (24%) | Drinking status | |
| 80+ years | 169 (3.9%) | Light drinker | 1,244 (45%) |
| Sex | | Heavy drinker | 370 (16%) |
| female | 1,465 (50%) | Moderate drinker | 207 (8.3%) |
| male | 1,471 (50%) | Non-drinker | 1,115 (31%) |
| Race | | Diabetes | |
| Non-Hispanic White | 1,022 (64%) | Yes | 379 (9.2%) |
| Non-Hispanic Black | 624 (10%) | No | 2,557 (90.8%) |
| Mexican American | 465 (9.1%) | Blood Cd | 0.26 (0.17, 0.46) |
| Other/multiracial | 484 (9.4%) | Urinary Cd | 0.18 (0.08, 0.33) |
| Other Hispanic | 341 (6.6%) | Ln (Blood Cd) | -1.35 (-1.77, -0.78) |
| Education attainment | | Ln (Urinary Cd) | -6.32 (-6.90, -5.69) |
| ≤9th | 281 (4.5%) | Age | 49 (34, 61) |
| 9-11th | 336 (8.2%) | BMI | 29 (25, 34) |
| High School | 707 (25%) | Waist | 101 (89, 112) |
| College or AA degree | 893 (32%) | HSCRP | 1.9 (0.8, 4.1) |
| College Graduate+ | 719 (31%) | Heart disease | |
| Marital status | | Yes | 310 (7.9%) |
| Married | 1,780 (64%) | No | 2,626 (92%) |
| Dissolved marriage | 627 (19%) | | |
| Unmarried | 529 (18%) | | |

[1] n (unweighted) (%); Median (IQR)

**Table S3.** Univariate Logistic Regression to Cardiovascular Disease

| Characteristics | OR | 95% CI | P-value | Characteristics | OR | 95% CI | P-value |
|---|---|---|---|---|---|---|---|
| Ln (Blood Cd) | 1.60 | 1.34, 1.92 | <0.001 | Marital status | | | |
| Blood Cd | | | | Married | Ref | — | |
| Q1 | — | — | | Unmarried | 0.58 | 0.28, 1.20 | 0.14 |
| Q2 | 1.96 | 0.98, 3.94 | 0.058 | Dissolved marriage | 2.57 | 1.82, 3.63 | <0.001 |
| Q3 | 2.86 | 1.50, 5.46 | 0.002 | Recreation activity | | | |
| Q4 | 3.72 | 2.30, 6.02 | <0.001 | No or lower | Ref | — | |
| Ln (Urinary Cd) | 2.23 | 1.86, 2.68 | <0.001 | Moderate | 0.88 | 0.55, 1.42 | 0.6 |
| Urinary Cd | | | | Vigorous | 0.35 | 0.19, 0.65 | 0.002 |
| Q1 | — | — | | Smoking status | | | |
| Q2 | 1.35 | 0.57, 3.21 | 0.5 | Never smoker | Ref | — | |
| Q3 | 2.77 | 1.43, 5.35 | 0.004 | Former smoker | 1.87 | 1.16, 3.02 | 0.012 |
| Q4 | 4.12 | 2.06, 8.22 | <0.001 | Current smoker | 1.74 | 1.09, 2.79 | 0.023 |
| Age | 1.08 | 1.07, 1.09 | <0.001 | Drinking status | | | |
| Grouped Age | | | | Non-drinker | Ref | — | |
| 20-39 | — | — | | Light drinker | 0.82 | 0.57, 1.18 | 0.3 |
| 40-59 | 4.81 | 2.32, 9.99 | <0.001 | Moderate drinker | 0.28 | 0.15, 0.53 | <0.001 |
| 60-79 | 18.8 | 9.01, 39.2 | <0.001 | Heavy drinker | 0.61 | 0.33, 1.11 | 0.1 |
| 80+ | 39.8 | 19.5, 80.9 | <0.001 | Diabetes | | | |
| Sex | | | | NO | — | — | |
| Female | — | — | | YES | 2.19 | 1.37, 3.50 | 0.002 |
| Male | 1.15 | 0.84, 1.58 | 0.4 | BMI group | | | |
| Race | | | | <18.5 | — | — | |
| Other/multiracial | — | — | | 18.5-25 | 0.81 | 0.24, 2.74 | 0.7 |
| Mexican American | 1.3 | 0.60, 2.82 | 0.5 | 25-30 | 0.92 | 0.31, 2.69 | 0.9 |
| Other Hispanic | 1.95 | 0.93, 4.09 | 0.076 | ≥30 | 1.06 | 0.35, 3.28 | >0.9 |
| Non-Hispanic White | 2.95 | 1.52, 5.73 | 0.003 | Waist | 1.02 | 1.01, 1.03 | 0.001 |
| Non-Hispanic Black | 3.14 | 1.74, 5.65 | <0.001 | HSCRP | 1.03 | 1.00, 1.05 | 0.027 |
| Education attainment | | | | | | | |
| ≤9th | — | — | | | | | |
| 9-11th | 1.41 | 0.78, 2.54 | 0.2 | | | | |
| High School | 1.06 | 0.54, 2.09 | 0.9 | | | | |
| College or AA degree | 0.83 | 0.43, 1.62 | 0.6 | | | | |
| College Graduate+ | 0.62 | 0.26, 1.47 | 0.3 | | | | |

**Adaptive LASSO analysis based on Former smoker *vs.* Never smoker samples**

**Table S4:** Adaptive LASSO of Variables Associated CVD

| Variable [value] | β | SE | Wald χ2 | P-value |
|---|---|---|---|---|
| Intercept | -8.7178 | 0.6564 | 176.3958 | <.0001* |
| Ln (Blood Cd) | 0.0000 | 0.0000 | 0.0000 | 1.0000 |
| Ln (Urinary Cd) | 0.0000 | 0.0000 | 0.0000 | 1.0000 |
| Smoking[Former smoker-Never smoker] | 0.0737 | 0.1537 | 0.2298 | 0.6316 |
| Waist circumference | 0.0179 | 0.0051 | 12.4299 | 0.0004* |
| Age | 0.0750 | 0.0059 | 159.8944 | <.0001* |
| Race[Mexican American-Other/multiracial] | 0.0000 | 0.0000 | 0.0000 | 1.0000 |
| Race[Other Hispanic-Other/multiracial] | 0.2212 | 0.2503 | 0.7815 | 0.3767 |
| Race[Non-Hispanic White-Other/multiracial] | 0.1941 | 0.1849 | 1.1020 | 0.2938 |
| Race[Non-Hispanic Black-Other/multiracial] | 0.3716 | 0.2160 | 2.9606 | 0.0853 |
| Marital status[Unmarried-Married] | 0.0000 | 0.0000 | 0.0000 | 1.0000 |
| Marital status[Dissolved marriage-Married] | 0.0000 | 0.0000 | 0.0000 | 1.0000 |
| Recreation activity[Moderate-No or lower] | 0.0000 | 0.0000 | 0.0000 | 1.0000 |
| Recreation activity[Vigorous-No or lower] | 0.0000 | 0.0000 | 0.0000 | 1.0000 |
| Drinking[Light drinker-Non-drinker] | 0.0000 | 0.0000 | 0.0000 | 1.0000 |
| Drinking[Moderate drinker-Non-drinker] | 0.0000 | 0.0000 | 0.0000 | 1.0000 |
| Drinking[Heavy drinker-Non-drinker] | 0.0000 | 0.0000 | 0.0000 | 1.0000 |
| Diabetes[YES-NO] | 0.0000 | 0.0000 | 0.0000 | 1.0000 |
| HSCRP | 0.0000 | 0.0000 | 0.0000 | 1.0000 |

**Note:** Covariates are included based on results of the univariate analysis in Table S3

**Table S5:** Adaptive LASSO of Variables Associated CVD

| Variable [value] | β | SE | Wald χ2 | P-value |
|---|---|---|---|---|
| Intercept | -0.7818 | 0.5070 | 2.3777 | 0.1231 |
| Ln (Blood Cd) | 0.0000 | 0.0000 | 0.0000 | 1.0000 |
| Ln (Urinary Cd) | 0.3694 | 0.0816 | 20.4839 | <.0001* |
| Smoking[Former smoker-Never smoker] | 0.4841 | 0.1474 | 10.7911 | 0.0010* |
| Race[Mexican American-Other/multiracial] | 0.0000 | 0.0000 | 0.0000 | 1.0000 |
| Race[Other Hispanic-Other/multiracial] | 0.6241 | 0.2473 | 6.3695 | 0.0116* |
| Race[Non-Hispanic White-Other/multiracial] | 1.0618 | 0.1870 | 32.2260 | <.0001* |
| Race[Non-Hispanic Black-Other/multiracial] | 0.8343 | 0.2134 | 15.2893 | <.0001* |
| Marital status[Unmarried-Married] | 0.0000 | 0.0000 | 0.0000 | 1.0000 |
| Marital status[Dissolved marriage-Married] | 0.3703 | 0.1619 | 5.2312 | 0.0222* |
| Recreation activity[Moderate-No or lower] | 0.0000 | 0.0000 | 0.0000 | 1.0000 |
| Recreation activity[Vigorous-No or lower] | -0.7309 | 0.1938 | 14.2193 | 0.0002* |
| Drinking[Light drinker-Non-drinker] | 0.0000 | 0.0000 | 0.0000 | 1.0000 |
| Drinking[Moderate drinker-Non-drinker] | 0.0000 | 0.0000 | 0.0000 | 1.0000 |
| Drinking[Heavy drinker-Non-drinker] | -0.1056 | 0.2022 | 0.2731 | 0.6013 |

| Variable [value] | β | SE | Wald χ2 | P-value |
|---|---|---|---|---|
| Diabetes[YES-NO] | 0.2964 | 0.1953 | 2.3046 | 0.1290 |
| HSCRP | 0.0000 | 0.0000 | 0.0000 | 1.0000 |

**Note:** Remove the first layer of variables, i.e., Waist and Age in Table S4

**Table S6:** Adaptive LASSO of Variables Associated CVD

| Variable [value] | β | SE | Wald χ2 | P-value |
|---|---|---|---|---|
| Intercept | -1.8939 | 0.1474 | 165.1360 | <.0001* |
| Ln (Blood Cd) | 0.2612 | 0.0961 | 7.3911 | 0.0066* |
| Drinking[Light drinker-Non-drinker] | 0.0000 | 0.0000 | 0.0000 | 1.0000 |
| Drinking[Moderate drinker-Non-drinker] | 0.0000 | 0.0000 | 0.0000 | 1.0000 |
| Drinking[Heavy drinker-Non-drinker] | 0.0000 | 0.0000 | 0.0000 | 1.0000 |
| Diabetes[YES-NO] | 0.3430 | 0.1973 | 3.0225 | 0.0821 |
| HSCRP | 0.0000 | 0.0000 | 0.0000 | 1.0000 |

**Note:** Remove the second layer of variables, i.e., Urinary Cd, Smoking, Race, Marital status and Recreation activity in Table S5

**Table S7:** Adaptive LASSO of Variables Associated CVD (Remove the third layer of variable)

| Variable [value] | β | SE | Wald χ2 | P-value |
|---|---|---|---|---|
| Intercept | -2.2149 | 0.0714 | 963.3795 | <.0001* |
| Drinking[Light drinker-Non-drinker] | 0.0000 | 0.0000 | 0.0000 | 1.0000 |
| Drinking[Moderate drinker-Non-drinker] | 0.0000 | 0.0000 | 0.0000 | 1.0000 |
| Drinking[Heavy drinker-Non-drinker] | 0.0000 | 0.0000 | 0.0000 | 1.0000 |
| Diabetes[YES-NO] | 0.2264 | 0.2099 | 1.1633 | 0.2808 |
| HSCRP | 0.0000 | 0.0000 | 0.0000 | 1.0000 |

**Note:** Remove the third layer of variable of Blood Cd in Table S6

**Adaptive LASSO analysis based on Current smoker *vs.* Never smoker samples**

**Table S8:** Adaptive LASSO of Variables Associated CVD

| Variable [value] | $\beta$ | SE | Wald $\chi 2$ | *P-value* |
|---|---|---|---|---|
| Intercept | -7.5635 | 0.6526 | 134.3145 | <.0001* |
| Ln (Blood Cd) | 0.0000 | 0.0000 | 0.0000 | 1.0000 |
| Ln (Urinary Cd) | 0.0000 | 0.0000 | 0.0000 | 1.0000 |
| Smoking[Current smoker-Never smoker] | 0.6822 | 0.1798 | 14.3976 | 0.0001* |
| Waist | 0.0106 | 0.0056 | 3.5378 | 0.0600 |
| Age | 0.0701 | 0.0056 | 156.6804 | <.0001* |
| Race[Mexican American-Other/multiracial] | 0.0000 | 0.0000 | 0.0000 | 1.0000 |
| Race[Other Hispanic-Other/multiracial] | 0.0000 | 0.0000 | 0.0000 | 1.0000 |
| Race[Non-Hispanic White-Other/multiracial] | 0.0211 | 0.1713 | 0.0152 | 0.9020 |
| Race[Non-Hispanic Black-Other/multiracial] | 0.0000 | 0.0000 | 0.0000 | 1.0000 |
| Marital status[Unmarried-Married] | 0.0000 | 0.0000 | 0.0000 | 1.0000 |
| Marital status[Dissolved marriage-Married] | 0.0000 | 0.0000 | 0.0000 | 1.0000 |
| Recreation activity[Moderate-No or lower] | 0.0000 | 0.0000 | 0.0000 | 1.0000 |
| Recreation activity[Vigorous-No or lower] | 0.0000 | 0.0000 | 0.0000 | 1.0000 |
| Drinking[Light drinker-Non-drinker] | 0.0000 | 0.0000 | 0.0000 | 1.0000 |
| Drinking[Moderate drinker-Non-drinker] | 0.0000 | 0.0000 | 0.0000 | 1.0000 |
| Drinking[Heavy drinker-Non-drinker] | 0.0000 | 0.0000 | 0.0000 | 1.0000 |
| Diabetes[YES-NO] | 0.0000 | 0.0000 | 0.0000 | 1.0000 |
| HSCRP | 0.0047 | 0.0130 | 0.1272 | 0.7214 |

**Note:** Covariates are included based on results of the univariate analysis in Table S3

**Table S9:** Adaptive LASSO of Variables Associated CVD

| Variable [value] | $\beta$ | SE | Wald $\chi 2$ | *P-value* |
|---|---|---|---|---|
| Intercept | -0.6842 | 0.6108 | 1.2548 | 0.2626 |
| Ln (Blood Cd) | 0.0000 | 0.0000 | 0.0000 | 1.0000 |
| Ln (Urinary Cd) | 0.5037 | 0.0854 | 34.8180 | <.0001* |
| Waist | 0.0110 | 0.0048 | 5.2179 | 0.0224* |
| Race[Mexican American-Other/multiracial] | 0.0000 | 0.0000 | 0.0000 | 1.0000 |
| Race[Other Hispanic-Other/multiracial] | 0.0000 | 0.0000 | 0.0000 | 1.0000 |
| Race[Non-Hispanic White-Other/multiracial] | 0.6037 | 0.1795 | 11.3158 | 0.0008* |
| Race[Non-Hispanic Black-Other/multiracial] | 0.2077 | 0.2158 | 0.9263 | 0.3358 |
| Marital status[Unmarried-Married] | 0.0000 | 0.0000 | 0.0000 | 1.0000 |
| Marital status[Dissolved marriage-Married] | 0.1461 | 0.1874 | 0.6083 | 0.4354 |
| Recreation activity[Moderate-No or lower] | 0.0000 | 0.0000 | 0.0000 | 1.0000 |
| Recreation activity[Vigorous-No or lower] | -0.6136 | 0.2201 | 7.7744 | 0.0053* |
| Drinking[Light drinker-Non-drinker] | 0.0000 | 0.0000 | 0.0000 | 1.0000 |
| Drinking[Moderate drinker-Non-drinker] | 0.0000 | 0.0000 | 0.0000 | 1.0000 |
| Drinking[Heavy drinker-Non-drinker] | 0.0000 | 0.0000 | 0.0000 | 1.0000 |

| Variable [value] | β | SE | Wald χ2 | P-value |
|---|---|---|---|---|
| Diabetes[YES-NO] | 0.3754 | 0.2268 | 2.7405 | 0.0978 |
| HSCRP | 0.0000 | 0.0000 | 0.0000 | 1.0000 |

**Note:** Remove the first layer of variables, i.e., Smoking and Age in Table S8

**Table S10:** Adaptive LASSO of Variables Associated CVD

| Variable [value] | β | SE | Wald χ2 | P-value |
|---|---|---|---|---|
| Intercept | -2.3202 | 0.1475 | 247.5693 | <.0001* |
| Ln (Blood Cd) | 0.2172 | 0.0845 | 6.6113 | 0.0101* |
| Marital status[Unmarried-Married] | -0.2937 | 0.1947 | 2.2766 | 0.1313 |
| Marital status[Dissolved marriage-Married] | 0.4117 | 0.1823 | 5.1005 | 0.0239* |
| Drinking[Light drinker-Non-drinker] | 0.0000 | 0.0000 | 0.0000 | 1.0000 |
| Drinking[Moderate drinker-Non-drinker] | 0.0000 | 0.0000 | 0.0000 | 1.0000 |
| Drinking[Heavy drinker-Non-drinker] | 0.0000 | 0.0000 | 0.0000 | 1.0000 |
| Diabetes[YES-NO] | 0.5312 | 0.2166 | 6.0130 | 0.0142* |
| HSCRP | 0.0135 | 0.0106 | 1.6279 | 0.2020 |

**Note:** Remove the second layer of variables, i.e., Urinary Cd, Waist, Race and Recreation activity in Table S9

**Table S11:** Adaptive LASSO of Variables Associated CVD

| Variable [value] | β | SE | Wald χ2 | P-value |
|---|---|---|---|---|
| Intercept | -2.3802 | 0.0941 | 640.2038 | <.0001* |
| Drinking[Light drinker-Non-drinker] | 0.0000 | 0.0000 | 0.0000 | 1.0000 |
| Drinking[Moderate drinker-Non-drinker] | 0.0000 | 0.0000 | 0.0000 | 1.0000 |
| Drinking[Heavy drinker-Non-drinker] | 0.0000 | 0.0000 | 0.0000 | 1.0000 |
| HSCRP | 0.0061 | 0.0161 | 0.1423 | 0.7060 |

**Note:** Remove the third layer of variable of Blood Cd, Marital status and Diabetes in Table S10

**Adaptive LASSO network analysis based on Former smoker *vs.* Never smoker samples**

**Table S12:** The relationship between the first-level variables and the second-level variables

| Outcome | Variable [value] | β | SE | Wald χ2 | P-value |
|---|---|---|---|---|---|
| Ln (Urinary Cd) | Intercept | -6.8026 | 0.1052 | 4178.9440 | <.0001* |
| | Age | 0.0233 | 0.0009 | 730.8341 | <.0001* |
| | Waist | -0.0061 | 0.0010 | 38.3107 | <.0001* |
| Former smoker | Intercept | -2.9767 | 0.2970 | 100.4213 | <.0001* |
| | Age | 0.0221 | 0.0026 | 69.9832 | <.0001* |
| | Waist | 0.0098 | 0.0027 | 13.0742 | 0.0003* |
| Race | Mexican American: Intercept | -3.8472 | 0.4332 | 78.8770 | <.0001* |
| | Mexican American: Age | 0.0000 | 0.0000 | 0.0000 | 1.0000 |
| | Mexican American: Waist | 0.0392 | 0.0044 | 78.7745 | <.0001* |
| | Other Hispanic: Intercept | -3.1988 | 0.4722 | 45.8897 | <.0001* |
| | Other Hispanic: Age | 0.0000 | 0.0000 | 0.0000 | 1.0000 |
| | Other Hispanic: Waist | 0.0302 | 0.0048 | 38.9223 | <.0001* |
| | Non-Hispanic White: Intercept | -4.5526 | 0.4375 | 108.2941 | <.0001* |
| | Non-Hispanic White: Age | 0.0183 | 0.0027 | 44.7768 | <.0001* |
| | Non-Hispanic White: Waist | 0.0435 | 0.0044 | 98.6999 | <.0001* |
| | Non-Hispanic Black: Intercept | -4.1469 | 0.4769 | 75.6176 | <.0001* |
| | Non-Hispanic Black: Age | 0.0000 | 0.0000 | 0.0000 | 1.0000 |
| | Non-Hispanic Black: Waist | 0.0435 | 0.0048 | 82.0099 | <.0001* |
| Marital status | Unmarried: Intercept | 1.1477 | 0.1827 | 39.4407 | <.0001* |
| | Unmarried: Age | -0.0562 | 0.0044 | 164.3001 | <.0001* |
| | Unmarried: Waist | 0.0000 | 0.0000 | 0.0000 | 1.0000 |
| | Dissolved marriage: Intercept | -3.1766 | 0.1952 | 264.9494 | <.0001* |
| | Dissolved marriage: Age | 0.0361 | 0.0032 | 123.7184 | <.0001* |
| | Dissolved marriage: Waist | 0.0000 | 0.0000 | 0.0000 | 1.0000 |
| Recreation activity | Moderate: Intercept | -0.7206 | 0.0509 | 200.5403 | <.0001* |
| | Moderate: Age | 0.0000 | 0.0000 | 0.0000 | 1.0000 |
| | Moderate: Waist | 0.0000 | 0.0000 | 0.0000 | 1.0000 |
| | Vigorous: Intercept | 2.6969 | 0.3051 | 78.1200 | <.0001* |
| | Vigorous: Age | -0.0365 | 0.0029 | 154.3103 | <.0001* |
| | Vigorous: Waist | -0.0155 | 0.0030 | 27.0632 | <.0001* |

**Table S13:** The relationship between the second-level variables and the third-level variables

| Outcome | Variable [value] | β | SE | Wald χ2 | P-value |
|---|---|---|---|---|---|
| Ln (Blood - Cd) | Intercept | 1.8518 | 0.0843 | 482.0827 | <.0001* |
| | Ln (Urinary Cd) | 0.5006 | 0.0134 | 1398.9412 | <.0001* |
| | Smoking[Former smoker-Never smoker] | 0.0859 | 0.0237 | 13.1579 | 0.0003* |
| | Race[Mexican American-Other/multiracial] | -0.1526 | 0.0301 | 25.6621 | <.0001* |
| | Race[Other Hispanic-Other/multiracial] | -0.2014 | 0.0341 | 34.9342 | <.0001* |
| | Race[Non-Hispanic White-Other/multiracial] | -0.1612 | 0.0261 | 38.1092 | <.0001* |
| | Race[Non-Hispanic Black-Other/multiracial] | 0.0000 | 0.0000 | 0.0000 | 1.0000 |
| | Marital status[Unmarried-Married] | 0.0000 | 0.0000 | 0.0000 | 1.0000 |
| | Marital status[Dissolved marriage-Married] | 0.0000 | 0.0000 | 0.0000 | 1.0000 |
| | Recreation activity[Moderate-No or lower] | -0.0305 | 0.0245 | 1.5443 | 0.2140 |
| | Recreation activity[Vigorous-No or lower] | 0.0000 | 0.0000 | 0.0000 | 1.0000 |

**Adaptive LASSO network analysis based on Current smoker *vs.* Never smoker samples**

**Table S14:** The relationship between the first-level variables and the second-level variables

| Outcome | Variable [value] | β | SE | Wald χ2 | P-value |
|---|---|---|---|---|---|
| Ln (Urinary Cd) | Intercept | -7.4848 | 0.0502 | 22273.0760 | <.0001* |
|  | Smoking[Current smoker-Never smoker] | 0.4547 | 0.0373 | 148.3490 | <.0001* |
|  | Age | 0.0244 | 0.0009 | 663.2454 | <.0001* |
| Waist | Intercept | 96.0052 | 1.1460 | 7018.6596 | <.0001* |
|  | Smoking[Current smoker-Never smoker] | 0.0000 | 0.0000 | 0.0000 | 1.0000 |
|  | Age | 0.0694 | 0.0211 | 10.8328 | 0.0010* |
| Race | Mexican American: Intercept | -0.1125 | 0.0724 | 2.4144 | 0.1202 |
|  | Mexican American: Smoking[Current smoker-Never smoker] | 0.0000 | 0.0000 | 0.0000 | 1.0000 |
|  | Mexican American: Age | 0.0000 | 0.0000 | 0.0000 | 1.0000 |
|  | Other Hispanic: Intercept | -0.5250 | 0.0816 | 41.3815 | <.0001* |
|  | Other Hispanic: Smoking[Current smoker-Never smoker] | 0.0000 | 0.0000 | 0.0000 | 1.0000 |
|  | Other Hispanic: Age | 0.0000 | 0.0000 | 0.0000 | 1.0000 |
|  | Non-Hispanic White: Intercept | -0.1089 | 0.1568 | 0.4826 | 0.4873 |
|  | Non-Hispanic White: Smoking[Current smoker-Never smoker] | 0.5867 | 0.1142 | 26.3848 | <.0001* |
|  | Non-Hispanic White: Age | 0.0107 | 0.0028 | 14.0545 | 0.0002* |
|  | Non-Hispanic Black: Intercept | 0.0565 | 0.0744 | 0.5766 | 0.4476 |
|  | Non-Hispanic Black: Smoking[Current smoker-Never smoker] | 0.6296 | 0.1251 | 25.3345 | <.0001* |
|  | Non-Hispanic Black: Age | 0.0000 | 0.0000 | 0.0000 | 1.0000 |
| Recreation activity | Moderate: Intercept | -0.7597 | 0.0530 | 205.6525 | <.0001* |
|  | Moderate: Smoking[Current smoker-Never smoker] | 0.0000 | 0.0000 | 0.0000 | 1.0000 |
|  | Moderate: Age | 0.0000 | 0.0000 | 0.0000 | 1.0000 |
|  | Vigorous: Intercept | 1.1665 | 0.1533 | 57.9101 | <.0001* |
|  | Vigorous: Smoking[Current smoker-Never smoker] | -0.4559 | 0.1156 | 15.5616 | <.0001* |
|  | Vigorous: Age | -0.0382 | 0.0032 | 141.9792 | <.0001* |

**Table S15:** The relationship between the second-level variables and the third-level variables

| Outcome | Variable [value] | β | SE | Wald χ2 | P-value |
|---|---|---|---|---|---|
| Ln (Blood Cd) | Intercept | 2.4447 | 0.1102 | 492.4839 | <.0001* |
| | Ln (Urinary Cd) | 0.5727 | 0.0180 | 1009.6926 | <.0001* |
| | Waist | 0.0000 | 0.0000 | 0.0000 | 1.0000 |
| | Race[Mexican American-Other/multiracial] | -0.0780 | 0.0397 | 3.8595 | 0.0495* |
| | Race[Other Hispanic-Other/multiracial] | -0.1239 | 0.0482 | 6.6110 | 0.0101* |
| | Race[Non-Hispanic White-Other/multiracial] | 0.0000 | 0.0000 | 0.0000 | 1.0000 |
| | Race[Non-Hispanic Black-Other/multiracial] | 0.1245 | 0.0407 | 9.3751 | 0.0022* |
| | Recreation activity[Moderate-No or lower] | 0.0000 | 0.0000 | 0.0000 | 1.0000 |
| | Recreation activity[Vigorous-No or lower] | 0.0000 | 0.0000 | 0.0000 | 1.0000 |
| Marital status | Unmarried: Intercept | -3.4652 | 0.5367 | 41.6941 | <.0001* |
| | Unmarried :Ln (Urinary Cd) | -0.5104 | 0.0677 | 56.7672 | <.0001* |
| | Unmarried: Waist | -0.0112 | 0.0036 | 9.5606 | 0.0020* |
| | Unmarried: Race[Mexican American-Other/multiracial] | 0.0000 | 0.0000 | 0.0000 | 1.0000 |
| | Unmarried: Race[Other Hispanic-Other/multiracial] | 0.0000 | 0.0000 | 0.0000 | 1.0000 |
| | Unmarried: Race[Non-Hispanic White-Other/multiracial] | 0.0000 | 0.0000 | 0.0000 | 1.0000 |
| | Unmarried: Race[Non-Hispanic Black-Other/multiracial] | 0.9869 | 0.1267 | 60.6547 | <.0001* |
| | Unmarried: Recreation activity[Moderate-No or lower] | 0.0000 | 0.0000 | 0.0000 | 1.0000 |
| | Unmarried: Recreation activity[Vigorous-No or lower] | 0.0000 | 0.0000 | 0.0000 | 1.0000 |
| | Dissolved marriage: Intercept | 1.3391 | 0.3874 | 11.9470 | 0.0005* |
| | Dissolved marriage: Ln (Urinary Cd) | 0.5184 | 0.0653 | 62.9788 | <.0001* |
| | Dissolved marriage: Waist | 0.0000 | 0.0000 | 0.0000 | 1.0000 |
| | Dissolved marriage: Race[Mexican American-Other/multiracial] | 0.2144 | 0.2092 | 1.0496 | 0.3056 |
| | Dissolved marriage: Race[Other Hispanic-Other/multiracial] | 0.6611 | 0.2189 | 9.1221 | 0.0025* |
| | Dissolved marriage: Race[Non-Hispanic White-Other/multiracial] | 0.9151 | 0.1599 | 32.7320 | <.0001* |
| | Dissolved marriage: Race[Non-Hispanic Black-Other/multiracial] | 1.2926 | 0.1730 | 55.8060 | <.0001* |
| | Dissolved marriage: Recreation activity[Moderate-No or lower] | 0.0000 | 0.0000 | 0.0000 | 1.0000 |
| | Dissolved marriage: Recreation activity[Vigorous-No or lower] | 0.0000 | 0.0000 | 0.0000 | 1.0000 |
| Diabetes | Ln (Urinary Cd) | 0.1386 | 0.0727 | 3.6407 | 0.0564 |
| | Waist | 0.0355 | 0.0036 | 99.0331 | <.0001* |
| | Race[Mexican American-Other/multiracial] | 0.1178 | 0.1957 | 0.3625 | 0.5471 |

| | | | | |
|---|---|---|---|---|
| Race[Other Hispanic-Other/multiracial] | 0.0000 | 0.0000 | 0.0000 | 1.0000 |
| Race[Non-Hispanic White-Other/multiracial] | -0.3604 | 0.1523 | 5.6009 | 0.0180* |
| Race[Non-Hispanic Black-Other/multiracial] | 0.0000 | 0.0000 | 0.0000 | 1.0000 |
| Recreation activity[Moderate-No or lower] | 0.0000 | 0.0000 | 0.0000 | 1.0000 |
| Recreation activity[Vigorous-No or lower] | -0.8727 | 0.1972 | 19.5909 | <.0001* |